\documentstyle[graphics,12pt,tighten,prb,aps]{revtex}
\newcommand{\kB}{k} 
\newcommand{\kT}{\kB T}
\newcommand{\lB}{l_{\mathrm{B}}}
\newcommand{\r}{{\mathbf{r}}} 
\newcommand{\dr}{d^3\r}
\newcommand{\nm}{\mathrm{nm}}
\newcommand{\um}{\mu\mathrm{m}}
\newcommand{\M}{\mathrm{M}}
\newcommand{\kappaT}{\kappa_{\mathrm{T}}}
\newcommand{\ro}{r_0}
\newcommand{\pot}{\psi}
\newcommand{\potD}{\overline{\pot}}
\newcommand{\smallplus}{{\scriptscriptstyle+}}
\newcommand{\smallminus}{{\scriptscriptstyle-}}
\newcommand{\smallpm}{{\scriptscriptstyle\pm}}
\newcommand{\np}{n_{\smallplus}}
\newcommand{\nn}{n_{\smallminus}}
\newcommand{\gpp}{g_{\smallplus\smallplus}}
\newcommand{\gpn}{g_{\smallplus\smallminus}}
\newcommand{\gnn}{g_{\smallminus\smallminus}}
\newcommand{\gxx}{g_{\smallpm\smallpm}}

\newcommand{\gmx}{g_{m\smallpm}}
\newcommand{\hpp}{h_{\smallplus\smallplus}}
\newcommand{\hpn}{h_{\smallplus\smallminus}}
\newcommand{\hnn}{h_{\smallminus\smallminus}}

\newcommand{\hmp}{h_{m\smallplus}}
\newcommand{\hmn}{h_{m\smallminus}}

\newcommand{\salt}{[\mathrm{salt}]}
\newcommand{\Zeff}{Z_{\mathrm{eff}}}
\newcommand{\Zmax}{Z_{\mathrm{max}}}
\newcommand{\Zmin}{Z_{\mathrm{min}}}
\newcommand{\Sbar}{\overline S}
\newcommand{\Scc}{S_{\mathrm{cc}}}
\newcommand{\Tstar}{T^*}
\newcommand{\DH}{Debye-H\"uckel} 
\newcommand{\WS}{Wigner-Seitz}
\newcommand{\PB}{Poisson-Boltzmann}
\newcommand{\PY}{Percus-Yevick}
\newcommand{\SI}{Sogami-Ise}
\newcommand{\eqref}[1]{(\ref{#1})}
\newcommand{\etal}{{\it et al}}
\newcommand{\loccit}{{\it loc. cit.}}

\newcommand{\role}{role}
\begin{document}
\title{A theory of void formation in charge-stabilised\\
colloidal suspensions at low ionic strength}
\author{Patrick B. Warren}
\address{Unilever Research Port Sunlight, 
Bebington, Wirral, CH63 3JW, UK.}
\date{JCP resubmission: \today}
\maketitle
\begin{abstract}
Using a carefully justified development of \DH\ theory for highly
asymmetric electrolytes, one finds that a region of expanded phase
instability, or miscibility gap, can appear for charge-stabilised
colloidal suspensions at high charges and low ionic strengths.  It is
argued that this is offers a straightforward explanation for the
observations of void structures and other anomalies in such
suspensions in this region.  The nature of the interface between
coexisting phases, and general arguments that many-body attractions
form a key part of the underlying physical picture, are also examined.
The present analysis may also generate new insights into old problems
such as coacervation in oppositely charged colloid or protein /
polyelectrolyte mixtures, and suggests interesting new possibilities
such as the appearance of charge density wave phases in colloidal
systems in the vicinity of the critical solution points.
\end{abstract}
\pacs{PACS: 05.20.-y, 64.75.+g, 82.70.Dd}

\section{Introduction}
For the past fifty years or so, our understanding of the stability of
lyophobic colloids has been guided by the seminal works of Derjaguin and
Landau,\cite{DL} and Vervey and Overbeek\cite{VO} (DLVO). The
focus is on pair potentials between colloid particles, which are
comprised of van der Waals attractions and electrostatic repulsions.
If the electrostatic repulsions are sufficiently large, the suspension
is stable.  On the other hand if the electrostatic repulsions are
reduced in some way, by screening by added electrolyte for example, a
suspension is rendered unstable towards flocculation.  The DLVO theory
gives a clear explanation for this `salting out' effect, first studied
in detail by Faraday.\cite{Faraday} Although the DLVO theory is aimed
squarely at understanding the kinetic stability of dilute colloidal
suspensions, the DLVO pair potentials are often used to rationalise many
other properties of colloidal suspensions.

The unquestioning use of the DLVO pair potentials has been challenged
over the last decade by experiments on dilute suspensions of highly
charged colloids at low ionic strength (where the pair potential would
predict absolute colloidal stability).  Void
structures,\cite{voids,YYKIH} vapour-liquid coexistence\cite{vapliq}
or biphasic regions,\cite{bi1,bi2} and other
anomalies\cite{USAXS,SSH2,SSH3,MIK} have been cited as evidence that
under these conditions an attractive minimum develops in the pair
potential at large distances, not captured by DLVO theory.  Several
reviews are available.\cite{paradox} Until recently, the only
theoretical explanation of the observed phenomena has been a
modification of the DLVO theory advanced by Sogami and Ise (SI),\cite{SI}
although their analysis has been challenged by other
workers.\cite{Woodward,KjMi}

Very recently though, van Roij, Dijkstra and Hansen
(RDH)\cite{vRH,vRDH} and Graf and L\"owen\cite{GL} have pointed out
the existence of so-called volume terms in the free energy of a
colloidal suspension, which can profoundly influence the stability of
such a suspension without affecting the pair potential of mean force
between colloid particles.  (The possible importance of volume terms
was first noted by Grimson and Silbert.\cite{GS}) The key idea is that
pair potentials are not the sole arbiters of phase stability, in the
presence of these volume terms.  Using these ideas, RDH found that
there could exist expanded regions of phase instability in the phase
diagram, at low volume fractions and ionic strengths, and argued that
this may provide an explanation of the anomalous phenomena.

Actually, as early as 1938, Langmuir had sounded a note of caution
about the pair potential approach.  He advanced three criticisms of
``the use of energy diagrams [pair potentials] to analyse the
stability of colloids.''\cite{Langmuir} These insights deserve to be
quoted verbatim from his paper:

``(A) No direct account is taken of the thermal agitation which by
itself would tend to cause the colloid particles and the ions to be
dispersed throughout the liquid giving an osmotic pressure $p=\sum nkT$.

``(B) The attraction between the charged micelles and the ion atmosphere
of opposite sign which extends throughout the intervening liquid is
ignored or neglected although it exceeds the repulsive force between
micelles.

``(C) The electric charges on the micelles are assumed to be constant,
whereas they must be, in general, dependent on the concentration of the
micelles.''\cite{Langmuir}

Langmuir argues that the electrostatic free energy, or more crucially
the electrostatic contribution to the osmotic compressibility, should
be \emph{negative}, corresponding to an effective \emph{attraction}.
This is because in an electrically neutral suspension,
correlations between unlike charges lower the free energy relative to
a state where all charges are distributed at random.  He supports this
argument with the physical examples of a salt crystal, and an
electrolyte solution. Further, he argues that to explain the
stability of charged colloidal suspensions (given that the van der
Waals forces are also attractive) one either requires ``some new kind
of \emph{repulsive} force'',\cite{Langmuir} or that the charge on the
surfaces of the colloid particles decreases sufficiently rapidly with
increasing concentration so as to make the suspension stable again.
The latter seems to be Langmuir's preferred explanation, hence point
(C) above.  He spends some time trying to set up a simple theory along
these lines using the ideas of Debye and H\"uckel\cite{DH} (DH) for simple
electrolytes, including a macroion contribution to the screening length.

Vervey and Overbeek raise several criticisms of Langmuir's approach,
pointing out that the DH linearisation approximation fails
severely, and also noting that ``[the double layer] thickness is
determined by the electrolyte concentration in the sol medium, far
from any particle, and is, therefore, independent of the sol
concentration (in dilute sols).''\cite{VO} In other words, the
macroions are too far apart to contribute to the screening length.
They conclude that ``Langmuir's theory of the attraction force between
particles is untenable.''\cite{VO} These criticisms are correct
(and below we shall have to address them in developing a new version
of the theory) but they also appear to have had the adverse effect of
driving Langmuir's general arguments into obscurity.\cite{RKS} In
fact, as I shall argue below, Langmuir's general points are well made.

The purpose of this paper therefore is to present a relatively simple
analysis of the free energy of a charge stabilised colloidal
suspension, treated as a highly asymmetric colloidal electrolyte, but
avoiding the problems associated Langmuir's original application of
\DH\ theory. In so doing we will find that the stability of charge
stabilised colloidal suspensions can be explained solely by the first
\emph{two} of Langmuir's points.  The principal factors are the
translational entropy of the small ions and finite macroion size
effects.  A more detailed examination though shows that these
stabilising mechanisms can \emph{fail}, at low salt, low volume
fractions, and high macroion charge.  This is the origin of the phase
instability discovered recently by RDH, which in the present model appears
as a \emph{closed-loop miscibility gap}---a region in the phase
diagram in which phase separation occurs into colloid rich and colloid
depleted phases.

The analysis is partly inspired by the recent application of DH
theory to the critical behaviour of the so-called `restricted
primitive model' (RPM) of symmetric electrolytes by Fisher and
co-workers,\cite{FL,Fisher} and a much earlier analysis of the
stability of parallel charged rods by Michaeli \etal.\cite{MOV} The
miscibility gap in the present case is the analogue of the
vapour-liquid phase transition found at low densities and low
temperatures in the RPM. There are many parallels between the present
approach to the colloid problem and the above cited approaches to the
RPM: for instance it is known that DH theory captures the phase
transition in the RPM provided the essential \role\ of finite ion size
effects is acknowledged,\cite{FL} and that the prediction of the
critical density can be inaccurate due to non-linear effects, largely
captured in the case of the RPM by the Bjerrum pairing model.\cite{DH}
These features will recur below.

The programme will be as follows.  First I specify a `primitive' model
of a colloidal suspension as a highly asymmetric electrolyte, by
analogy to the RPM.  Next I analyse the domains of applicability of
various approximations, before obtaining a tractable analytic
expression for the free energy.  As well as explaining the stability
of colloidal suspensions, this free energy also predicts the
closed-loop miscibility gaps at low ionic strength and high charge,
which are analysed in some detail.  The nature of interface between
coexisting phases is also examined, as are arguments for the existence
of many-body attractions. 

As well as presenting this theory in some detail, I will also try to
relate the ideas to the existing body of work which is quite
extensive.  In particular, in appendices, I present comparisons with
the \SI\ theory,\cite{SI} the recent work by van Roij, Dijkstra and
Hansen,\cite{vRH,vRDH} and that by Levin, Barbosa and
Tamashiro.\cite{LBT,LBT2}

\section{A primitive model for colloidal suspensions}
Suppose that the colloid particles or macroions have diameter
$\sigma=2a$, positive charge $Z\gg1$ and number density $n_m$.  The
macroion volume fraction is $\phi=4\pi a^3n_m/3$.  There are small
coions and counterions at number densities $\np$ and $\nn$
respectively.  I suppose there is only one species of counterion and
all small ions are univalent, and of a size sufficiently small to be
negligible in the analysis which follows.\cite{size}  The solvent is
taken to be a dielectric continuum of permittivity $\epsilon$.  The
Coulomb interaction between a pair of univalent charges in units of
$\kT$ is $\lB/r$ where $\lB=e^2/4\pi\epsilon\kT$ is the Bjerrum length
($0.72\,\nm$ in water at room temperature)---this is one of the
natural length scales for the problem.  The effects of temperature can
be subsumed into the weak temperature dependence of $\lB$.\cite{CPAH}
The densities are slaved by the electroneutrality condition $Z
n_m+\np=\nn$, which actually plays a rather important \role\ as
discussed further below.  It will be convenient to write $\np=n_s$ to
bring out the fact that it represents the added salt or electrolyte
concentration.

Much work has been done applying integral equation methods to this
model. The mean spherical approximation (MSA) can be solved
analytically,\cite{MSA} and it has been noticed that this predicts a
region of phase instability for the salt-free system.\cite{GrootMSA}
Similarly, solutions for the hypernetted chain (HNC) approximation
have been computed,\cite{HNC} and Belloni and others also found
indications of a phase instability.\cite{Belloni} Various improved
approximations have also been studied.\cite{IE} Other workers have
considered the problem using the random phase approximation
(RPA),\cite{RPA} the symmetrised \PB\ equation,\cite{SymmPB} an
analogue of the Fisher-Levin approach to the RPM,\cite{LBT,LBT2} various
local density approximations,\cite{StRo,varlda,TATKB} and most
recently field-theoretical methods.\cite{NO} Because of the difficulty
in extracting solutions though, a full phase diagram does not appear
to have been constructed. In addition, the physics underlying the
various approximations is often obscure, and it may not be easy to
distinguish numerical artifacts from real effects.

Moreover, in the integral equation approaches to the primitive model,
it may not be sensible to treat the macroion-macroion interactions on
the same footing as the macroion-small ion or small ion-small ion
interactions because of the gross asymmetry.  This is why it can be
desirable to integrate out the small ion degrees of freedom and treat
what remains as an effective one-component system with a different
approximation scheme.  The effective one component model has traps for
the unwary though: we shall see below that three of the most important
contributions to the overall free energy are volume terms, and as such
easily omitted.  (These are the background electrolyte free energy,
the interaction between the small ions and the macroions, and most
subtle of all, a background subtraction counterterm that cancels most
of the mean field macroion-macroion interaction.)

Very recently, direct numerical simulation of one particular realisation
of this model has been undertaken by Lobaskin and Linse,\cite{LoLi}
despite the difficulty in accessing relevent system sizes and time
scales. This work confirms Langmuir's insight that the pressure in these
systems is reduced by the electrostatic interactions.

To describe the physics of the model, it will be convenient to draw
analogies with the physics of classical plasmas\cite{SDS,Stishov} and
liquid metals.\cite{EH,Ash,SA}  Other authors, particularly Grimson and
co-workers,\cite{GS,GrOCP} have recognised the benefits of these
analogies.

\subsection{Domains of behaviour of the model}
My first aim is to establish the domain of applicability of the
various linearisation and mean-field approximations underpinning \DH\
theory. To this end I introduce the following two screening lengths:
$\kappa^{-1}$ which is the Debye screening length from the small ions,
and $\kappaT^{-1}$ which is the screening length assuming the
macroions also contribute. It seems something of a heresy to introduce
$\kappaT$ but the analysis below will indicate precisely the limits of
validity of this concept. The two screening lengths are given by
$\kappa^2=4\pi\lB(\np+\nn)=4\pi\lB(Zn_m+2n_s)$, and
$\kappaT^2=4\pi\lB(Z^2n_m+\np+\nn)\approx4\pi\lB(Z^2n_m+2n_s)$. It is
also convenient to introduce the \WS\ cell radius $\ro$ as a measure
of the typical distance between macroions, defined by $4\pi
\ro^3n_m/3=1$. Noting that $\kappaT^{-1}<\kappa^{-1}$ finds three
distinct regions of behavior depending on the relative magnitudes of
$\kappa^{-1}$, $\kappaT^{-1}$ and $\ro$: (I) $\kappaT\ro<1$, (II)
$\kappa\ro<1<\kappaT\ro$ and (III) $1<\kappa\ro$.

If there is no added salt, then $\kappaT^{-1}\sim Z^{-1}n_m^{-1/2}$,
$\kappa^{-1}\sim Z^{-1/2}n_m^{-1/2}$, and $\ro\sim n_m^{-1/3}$.  As
the macroion concentration increases therefore, the two screening
lengths decrease \emph{more} rapidly than the mean distance between
macroions. Thus region I is obtained at the lowest macroion
concentration, followed by region II and then region III. The
crossovers occur at $\lB^3n_m\sim Z^{-6}$ and $\lB^3n_m\sim Z^{-3}$
respectively (restoring factors of $\lB$).

\begin{figure}[t]
\begin{center}
\vspace{15pt}
\includegraphics{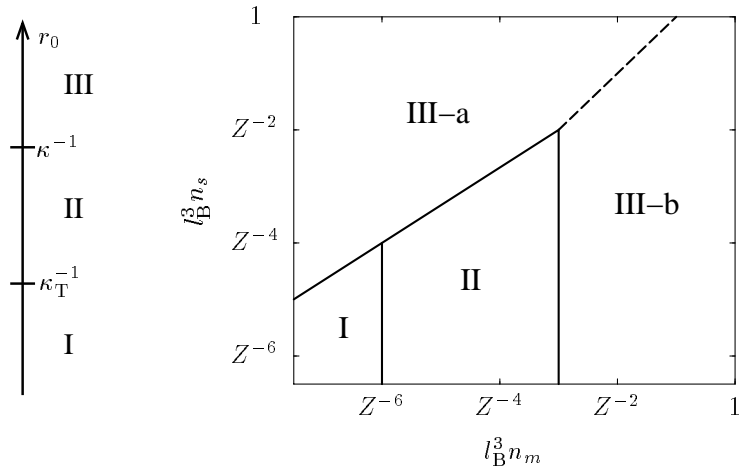}
\end{center}
\caption[?]{Generic classes of behaviour of a charge stabilised
colloidal suspension, treated as a highly asymmetric electrolyte: (I)
weakly coupled OCP, (II) strongly coupled OCP, (III) screened Coulomb
interaction divided into (III-a) added electrolyte dominates screening
and (III-b) colloid counterions dominate screening.\label{regfig}}
\end{figure}

If salt is added, it acts to reduce both screening lengths.  For
instance one has a crossover from $\kappa\sim Z^{1/2} n_m^{1/2}$ for
$n_s\ll Zn_m$ to $\kappa\sim n_s^{1/2}$ for $n_s\gg Zn_m$, and there
is a similar crossover for $\kappaT$.  This competes with the
crossovers discussed in the preceeding paragraph but when the dust
settles a fairly simple picture emerges, shown in Fig.~\ref{regfig}.
If $Z\gg1$ the crossovers are well separated. Note that the
analysis in the strict absence of salt fails to capture the re-entrant
behaviour of region III.
 
What is the physics behind this classification?  In region I there are
sufficient macroions in a screening volume for a DH mean-field
approximation to be correct, since $\kappaT r_0\ll1$ implies
$\kappaT^{-3}n_m\gg1$.  Below we shall see that the DH linearisation
approximation is also valid.  In region I therefore, Langmuir's
picture of DH theory applied to the system \emph{as a whole} is
valid.  In regions II and III though, the spacing between macroions is
much greater than the associated screening length
($\ro\gg\kappaT^{-1}$), and in fact both of the approximations
underlying Langmuir's picture break down.  (Thus it is only in region
I that $\kappaT$ has any validity.)  Note that this explanation of the
ineffectiveness of the macroions in contributing to the screening
length in regions II and III is not dependent on any time scale
separation between the dynamics of the macroions and the small ions
(Born-Oppenheimer or adiabatic approximation).\cite{SI,LBT}  Indeed, I
would argue this dynamic effect should play no \role\ in determining
the equilibrium properties of a colloidal suspension, but I shall
return to this point below.

In both regions I and II, the macroions are interacting with
effectively an \emph{unscreened} Coulomb law, $Z^2\lB/r$, since the
small ion screening length is still much greater than their mean
separation ($\kappa^{-1}\gg\ro$).  In these regions it is reasonable
to treat the macroions as a \emph{one-component plasma} (OCP) in a
neutralising background of electrolyte solution, where the
\emph{polarisation} of this backround by the macroions is a weak
perturbation.  On entering region III though, the screening length
from the small ions becomes less than or comparable to the distance
between neighbouring macroions, resulting in a strongly screened
macroion interaction.  The polarisation of the background electrolyte
is no longer weak and the macroions become surrounded with a
\emph{double layer} of small ions. Here it makes sense to follow
Verwey and Overbeek, and treat the macroions and their accompanying
double layers as composite objects.  In particular, Verwey and
Overbeek show that the original Coulomb interaction between macroions
is replaced by the well known DLVO screened Coulomb interaction
$Z^2\lB e^{-\kappa r}/r$, possibly with $Z$ replaced by
$\Zeff=Ze^{\kappa a}/(1+\kappa a)$.\cite{VO}

Let us return to the linearisation approximation.  In regions I and
II, I have argued above that one can consider the macroions as an
OCP. Since the macroion charge is $Z$, the corresponding plasma
coupling constant is $\Gamma=Z^2\lB/\ro\sim Z^2\lB n_m^{1/3}$.  Thus
the crossover from weak to strong coupling occurs at $\Gamma\sim1$ or
$\lB^3n_m\sim Z^{-6}$, which is precisely where the crossover between
regions I and II is found. Thus region I corresponds to an OCP in the
weak coupling regime where the DH linearisation approximation is
valid, whereas region II corresponds to the strong coupling regime
where correlation effects are important. In fact in region II one
would expect a freezing transition (colloidal crystals) at
$\Gamma\sim180$.\cite{SDS} The crossover from II to III occurs at
$\Gamma\sim Z$ so this transition should be accessible. The width of
the transition is determined by the free energy of the background in a
manner discussed recently by Stishov.\cite{Stishov}

The situation is summarised in Fig.~\ref{regfig}. In regions I and II,
the macroions are well approximated by an OCP in a neutralising and
weakly polarisable background. Region I (II) corresponds to the weak
(strong) coupling regime.  In region III, the polarisation of the
background becomes severe, and the Coulomb law is screened on
distances less than the mean spacing between macroions.

Let us put some typical numbers into the problem, for example $Z=1000$
and $\sigma=200\,\nm$. The upper apex of region II corresponds to
$\phi\sim10^{-2}$ and $\salt\sim10^{-6}\,\M$, and the upper apex of
region I corresponds to $\phi\sim10^{-11}$ and
$\salt\sim10^{-12}\,\M$.  These results are typical.  We see that
region I occurs at physically inaccessible concentrations, and in this
sense Verwey and Overbeek's criticism of Langmuir is made very
precise.  Region II is accessible at very low salt concentrations, and
is where the anomalous behavior is observed.  Most colloidal
suspensions sit in region III though, where the DLVO picture is expected
to be valid.

In the next section therefore I will develop a closed form expression
for the free energy appropriate to regions II and III.  \DH\
approximations will be used to handle the small ion-small ion and
small ion-macroion interactions, but, as the above analysis shows, the
macroions may be in a strong coupling regime.  For instance the macroion
pair correlation function may develop strong oscillations, which can
never be reproduced in a simple DH theory.  To handle the
macroion-macroion interactions therefore, I will turn to a variational
method first introduced by Firey and Ashcroft,\cite{FA} and applied to
colloidal suspensions by Shih and Stroud.\cite{SS}

\begin{figure}[t]
\begin{center}
\vspace{15pt}
\includegraphics{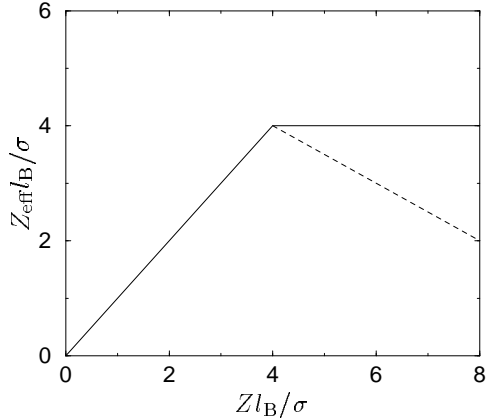}
\end{center}
\caption[?]{Schematic dependence of effective charge on a macroion as
the bare charge on a macroion is increased.  Above a certain point,
counterion condensation effectively takes place. This point is
$\Zmax\approx\alpha\sigma/\lB$ where $\alpha$ is a prefactor of order
unity.  For the purposes of the present work, $\alpha\approx4$ is
assumed, to correspond with previous work.\cite{LBT,ACGMPH,GrootIon}.
When fluctuation and correlation effects are taken into account, the
situation may correspond more closely to the dashed line (see Fig.~7
in Ref.~\onlinecite{GrootIon}).\label{grootfig}}
\end{figure}

Before moving on to this, I should discuss the approximation involved
in treating the small ion-macroion interaction (the double layer)
using DH theory.  In this case, the DH approximation amounts to
linearising $e^{-Z\lB/r}$ in the space arond the macroion. This should
be valid provided $Z\lB/r<1$ for $r>a=\sigma/2$, or
$Z\lB/\sigma<\alpha$ where $\alpha$ is a numerical constant of order
unity.  If $Z\lB/\sigma>\alpha$, the effective charge on the macroion
(as gauged by comparing with the DH result at large distances for
example) saturates around $\Zmax\approx\alpha\sigma/\lB$ by a kind of
counterion condensation effect.  This effect gives an indication of
the limit of validity of DH theory applied to the problem.  Using
Poisson-Boltzmann (PB) theory, Alexander \etal\ suggest that the
maximum charge is of order $15a/\lB$ suggesting that $\alpha\approx7$
(Eq. (I.7) in Ref.~\onlinecite{ACGMPH}). Simulations by Groot though
suggest that PB theory overestimates the effective charge, and
correlation effects may actually reduce the effective charge for
$Z>\Zmax$.\cite{GrootIon} Results resembling PB theory are also
obtained in a variational approach by Levin \etal.\cite{LBT} The
general phenomenon is illustrated in Fig.~\ref{grootfig}, where I have
compromised on $\alpha\approx4$ as representative of the literature as
a whole.  Note that the more conservative criterion, $\alpha\approx1$,
ignores the fact that most of the double layer lies at a distance of
order $\kappa^{-1}$ from the macroion surface (see below); a point
discussed in early work by Hartley.\cite{Hartley} Experimentally, the
effect is broadly confirmed in a number of studies,\cite{exptcond} and
very recently for one $Z$ and $\sigma$ it has been checked in direct
numerical simulations.\cite{LoLi}

\subsection{Analytic expression for the free energy}
To get a tractable expression for the free energy of the system, I
follow the development in Landau and Lifshitz,\cite{LL} and start with
an exact expression for the electrostatic energy in terms of the pair
correlation functions $g_{ij}(r)$
\begin{equation}
\frac{E}{V\kT}=\frac{1}{2}\sum_{i,j}n_in_j
\int\dr\,\frac{z_iz_j\lB}{r}\,g_{ij}(r)
\end{equation}
where $z_i=Z,1,-1$ as $i,j$ range over $m,+,-$, and $V$ is the volume
of the system.  We can replace $g_{ij}$ by $h_{ij}=g_{ij}-1$ in the
this expression by virtue of the electroneutrality condition, and
split the energy into $E = E_{\text{ss}} + E_{\text{ms}} +
E_{\text{mm}}$ where\cite{Exist}
\begin{eqnarray}
\frac{E_{\text{ss}}}{V\kT}&&=\frac{1}{2}\int\dr\,\frac{\lB}{r}\,
[\np^2\hpp(r)+\nn^2\hnn(r)-2\np\nn\hpn(r)],\label{esseq}\\
\frac{E_{\text{ms}}}{V\kT}&&=n_m\int\dr\,\frac{Z\lB}{r}\,
[\np\hmp(r)-\nn\hmn(r)],\\
\frac{E_{\text{mm}}}{V\kT}&&=\frac{n_m^2}{2}\int\dr\,\frac{Z^2\lB}{r}\,
h_{mm}(r).\label{emmeq}
\end{eqnarray}
To calculate these requires expressions for the $g_{ij}$.  As
discussed above, with care we can use DH theory to derive $\gxx$ and
$\gmx$.  The DH approximations are inapplicable for $g_{mm}$, which
can exhibit strong structural features in the regimes of interest.
The macroion contribution will therefore be handled separately by the
aforementioned variational method.

I now proceed by analogy with the DH theory for the RPM,\cite{DH,LL}
taking into account the analysis of the preceeding section.  We
solve for the electrostatic potential around an ion of charge $z$ by
solving the linearised \PB\ equation $\nabla^2\pot + \kappa^2\pot = 0$
in the space around the ion, where $\kappa^2 = 4\pi\lB(Zn_m+2n_s)$ is
from the \emph{small ions only} (regions II and III), and $\pot$ is
the electrostatic potential.  Next the potential of mean force between
this ion and another ion of charge $z'$ is taken to be $z'\pot$ so
that the pair correlation function between the ions is given by
$g(r)=e^{-z'\pot/\kT}\approx 1-z'\pot(r)/\kT$, applying the
linearisation approximation again.

For the small ions this leads to $\gpp=\gnn=1-\lB e^{-\kappa r}/r$ and 
$\gpn=1+\lB e^{-\kappa r}/r$ where the size of the ions has been
neglected. In the energy integral above this results in
\begin{equation}
\frac{E_{\text{ss}}}{V\kT}=-\frac{2\pi\lB^2(\np+\nn)^2}{\kappa}=
-\frac{\kappa^3}{8\pi}
\end{equation}
which is recognised as the DH internal energy in
a simple electrolyte solution.\cite{DH,LL}

For small ions around a large ion, the finite size of the macroion
should be taken into account.  Solving the linearised \PB\ equation
with the condition that the electric field at the surface matches the
surface charge density ($\partial\pot/\partial r=Z/4\pi a^2$ at 
$r=a$) results in an electrostatic potential
\begin{equation}
\pot(r)=\frac{Z\lB\kT}{r}\frac{e^{-\kappa(r-a)}}{1+\kappa a}.
\label{poteq}
\end{equation}
Writing $\gmx(r)=1\mp\pot(r)/\kT$ in the above energy integral gives
\begin{equation}
\frac{E_{\text{ms}}}{V\kT}=-\frac{Z^2\lB^2n_m(\np+\nn)}{1+\kappa a}
\int_a^\infty\!4\pi r^2dr\,\frac{e^{-\kappa(r-a)}}{r^2}
=-\frac{Z^2\lB\kappa n_m}{1+\kappa a}\label{emseq}
\end{equation}
which is recognised as the energy of a finite sized macroion in DH
theory.\cite{DH} Note that the integral has been truncated at $r=a$
since the hard core repulsion should not contribute to the internal
energy (it will be accounted for later).  Per macroion, this energy
takes the form $-Z^2\lB\kT/(a+\kappa^{-1})$ which has a well known
interpretation---it is the electrostatic energy between the macroion
and a counterion shell of equal and opposite charge situated a
distance $\kappa^{-1}$ away from the surface.\cite{Hartley}

\begin{figure}
\begin{center}
\vspace{15pt}
\includegraphics{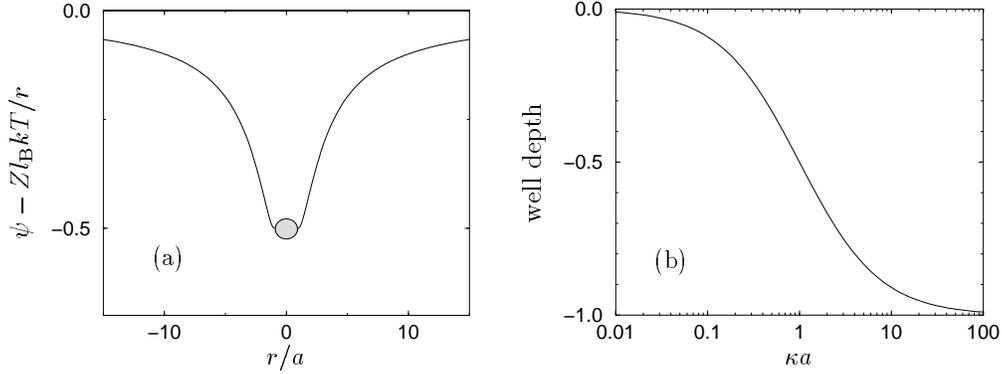}
\end{center}
\caption[?]{(a) Electrostatic potential field around a macroion due to
the double layer of small ions, at $\kappa a=1$.  The macroion
polarises the surrounding electrolyte and sits in a potential well of
its own making.  (b) Depth of well as a function of $\kappa a$.  In
both plots, the units of potential are $Z\lB\kT/a$.\label{potfig}}
\end{figure}

This result can also be obtained directly from the potential in
Eq.~\eqref{poteq} by recognising that $\pot(r)-Z\lB\kT/r$ is the
potential field due to the small ions around the macroion.\cite{LL}
This is illustrated in Fig.~\ref{potfig}(a), which clearly shows that
each macroion sits in a potential well due to the double layer of
small ions.  The interaction energy between the macroion and its
double layer is given by
\begin{equation}
Z\pot(a)-\frac{Z^2\lB\kT}{a}=
-\frac{Z^2\lB\kT\kappa}{1+\kappa a}\label{polengeq}.
\end{equation}
Multiplying by the number of macroions recovers Eq.~\eqref{emseq}.  In
plasma language, Eq.~\eqref{polengeq} is the DH estimate of the
\emph{polarisation energy} when a macroion is placed in an electrolyte
solution.  The depth of this well is shown as a function of $\kappa a$
in Fig.~\ref{potfig}(b). Note that it increases with increasing ionic
strength (increasing $\kappa$).  This means that there is a tendency
for macroions to drift towards areas of enhanced ionic strength.

The free energy contributions corresponding to $E_{\text{ss}}$ and
$E_{\text{ms}}$ are most readily obtained by the Debye charging
procedure,\cite{DH} which in the present case takes the form
\begin{equation}
\frac{F}{V\kT}=\int_0^1\frac{d\lambda}{\lambda}
\biggr(\frac{E}{V\kT}\biggl)_{\lB\to\lambda\lB}.
\end{equation}
Thus one obtains
\begin{equation}
\frac{F_{\text{ss}}}{V\kT}=-\frac{\kappa^3}{12\pi},\qquad
\frac{F_{\text{ms}}}{V\kT}=-\frac{2}{3}Z^2\lB\kappa n_m\,
f(\kappa a),\label{fsmseq}
\end{equation}
where the finite macroion size correction factor is
$f(x)=({3}/{x^3})[\log(1+x)-x+{x^2}/{2}]$ (note that $f(x)\to1$ as
$x\to0$).  The behaviour of $F_{\text{ms}}$ as a function of $\kappa
a$ is similar to $E_{\text{ms}}$.

Now let us return to the macroion-macroion contribution.  The energy
integral Eq.~\eqref{emmeq} is still correct, except that we can no
longer use DH theory for $g_{mm}(r)$ and are forced to turn to
another approach.  Before doing so, first note that we might
reasonably approximate $Z^2\lB/r$ with $Z^2\lB e^{-\kappa r}/r$ in
Eq.~\eqref{emmeq}.  This is because we would expect $h_{mm}\approx0$
for $r>\kappa^{-1}$, which is the region where the approximation is
most severe.  Then we can write
\begin{equation}
\frac{E_{\text{mm}}}{V\kT}\approx\frac{n_m^2}{2}
\int\dr\,\frac{Z^2\lB e^{-\kappa r}}{r}\,g_{mm}(r)
-\frac{2\pi Z^2\lB n_m^2}{\kappa^2}.\label{emm2eq}
\end{equation}
We recognise the first term in this as the internal energy of a system
consisting solely of macroions interacting with the potential of mean
force $Z^2\lB\kT e^{-\kappa r}/r$, since in the true system
$g_{mm}(r)$ does correspond to this screened Coulomb potential.  Such
a system is well defined even in the absence of oppositely charged
counterions.  The second term is a reminder that in the real system,
we should correct for the uniform neutralising background by
subtracting off the mean field ($g_{mm}=1$) term.  The present
analysis shows quite clearly that such a term exists; another proof
is described by RDH.\cite{vRDH}

Rather than proceeding further with the internal energy expression, we
will instead use the variational approach of Firey and
Ashcroft,\cite{FA} to obtain the free energy directly.  This approach
replaces the free energy by the variational minimum of $F_0+\langle
U\rangle_0$ where the subscript `0' in this case denotes a reference
system of hard spheres whose diameter (or volume fraction at fixed
number density) is the variational parameter.  In this method the
macroion-macroion contribution to the free energy is
\begin{equation}
\frac{F_{\text{mm}}}{V\kT}=n_m\frac{\eta(4-3\eta)}{(1-\eta)^2}
+\frac{2\pi Z^2\lB n_m^2}{\kappa^2}[\lambda^2G(\lambda)-1].\label{fmmeq}
\end{equation}
The first term is the free energy of the hard sphere
reference state, and the second is $\langle U\rangle_0$ evaluated in
closed form using the \PY\ pair correlation function for hard
spheres. Note that the background subtraction counterterm is
incorporated in this.\cite{SS}  Again, one may wish to use
$\Zeff=Ze^{\kappa a}/(1+\kappa a)$ instead of $Z$,\cite{VO} in which
case a factor $e^{2\kappa a}/(1+\kappa a)^2$ should be inserted in
front of $\lambda^2G(\lambda)$.  However it is important to note that
the background subtraction counterterm should \emph{not} be similarly
corrected.\cite{zeff}

In Eq.~\eqref{fmmeq} the variational parameter is the effective volume
fraction $\eta$, a parameter $\lambda=2\kappa\ro\eta^{1/3}$ is
introduced, and the following functions are defined:\cite{FA,SS}
\begin{eqnarray}
G(\lambda)&&={\lambda L(\lambda)}/
[{12\eta(L(\lambda)+\Sbar(\lambda)e^\lambda)}],\\
L(\lambda)&&=12\eta[(1+\eta/2)\lambda+(1+2\eta)],\\
\Sbar(\lambda)&&=(1-\eta)^2\lambda^3+6\eta(1-\eta)\lambda^2
+18\eta^2\lambda-12\eta(1+2\eta).
\end{eqnarray}
The variational minimum is found by solving the equation $\partial
F_{\text{mm}}/\partial\eta=0$ which has to be done numerically.

The final pieces in the free energy are the ideal gas or translational
entropy terms:
\begin{equation}
\frac{F_{\text{id,hc}}}{V\kT}=n_m\log n_m
+\np\log \frac{\np}{1-\phi}
+\nn\log \frac{\nn}{1-\phi}
\end{equation}
where the previously omitted small ion-macroion hard core repulsion is
captured by inserting correction factors of $1/(1-\phi)$.  We should
not make a similar correction for the macroion hard core interactions
since it is already in the hard sphere reference free energy in
Eq.~\eqref{fmmeq}.  It is convenient to split this into the proper
ideal term and a hard-core excluded volume term, and write (replacing
$\np$ and $\nn$ by $n_s$ and $(Zn_m+n_s)$ respectively)
\begin{eqnarray}
\frac{F_{\text{id}}}{V\kT}&&=n_m\log n_m+n_s\log n_s
+(Zn_m+n_s)\log(Zn_m+n_s),\label{fideq}\\
\frac{F_{\text{hc}}}{V\kT}&&=(Zn_m+2n_s)\log\frac{1}{1-\phi}.\label{fhceq}
\end{eqnarray}
We now have a closed form analytic free energy which consists of five
terms: 
\begin{equation}
F = F_{\text{id}} + F_{\text{hc}} + F_{\text{ss}} +
F_{\text{ms}} + F_{\text{mm}},\label{ftoteq}
\end{equation}
given in Eqs.~\eqref{fideq}, \eqref{fhceq}, \eqref{fsmseq}, and
\eqref{fmmeq} respectively (plus associated definitions). Note that
this final result is very close in spirit to the approach of Fisher
and Levin to the free energy of the RPM of simple
electrolytes.\cite{FL}  To summarise, the five terms in Eq.~\eqref{ftoteq} are
respectively: an ideal solution term, a small ion-macroion hard core
exclusion term, the small ion electrostatic free energy identical to
the DH free energy of the background electrolyte, the small
ion-macroion interaction free energy or background electrolyte
polarisation energy, and the macroion-macroion
interaction free energy which incorporates the background subtraction
counterterm.

Before discussing the consequential phase behaviour, let me recap the
nature of the approximations underlying this free energy. Firstly, it
is useful to contrast it with Langmuir's attempt to use DH theory for
the same problem. In the present theory, DH type approximations
involve the small ions only, and are applied to the small ion excess
free energy and the small ion-macroion interaction energy or
polarisation energy. The macroion-macroion interactions are treated
separately by an established variational procedure,\cite{FA,SS} and
tied in to the other contributions by the exact split of the internal
energy in Eqs.~\eqref{esseq}--\eqref{emmeq}.

The analysis is consistent with the preceeding section which shows
that in regions II and III only the small ions contribute to the
screening length, and that macroion-macroion interactions are either
strongly screened (region III) or in a strong coupling limit (region
II).  Additionally, the use of DH theory for the small ion-macroion
contribution is valid provided $Z\lB/\sigma$ is not too large.

It might appear that the variational approach to the macroion
interactions should break down in region II ($\kappa^{-1}\gg\ro$)
where one might expect $\eta>1$, supposing that the variational
macroion radius is of order $\kappa^{-1}$.  In fact the variational
estimate is good for this region too as can be seen from the following
argument.  In region II the macroions are essentially an OCP,
corresponding to the limit $\kappa\to0$ in Eq.~\eqref{fmmeq}, which
reduces to\cite{Ash,SA}
\begin{equation}
\frac{F_{\text{mm}}}{N_m\kT}=\eta\frac{(4-3\eta)}{(1-\eta)^2}
-3\Gamma\,\eta^{2/3}\frac{1-\eta/5+\eta^2/10}{1+2\eta}
\end{equation}
where $\Gamma=Z^2\lB/\ro$ is the plasma coupling constant introduced
earlier and $N_m=Vn_m$ is the number of macroions.  The variational
minimum gives the following implicit equation for $\eta$
\begin{equation}
\Gamma=2\eta^{1/3}\frac{(1+2\eta)^2(2-\eta)}{(1-\eta)^5(2+\eta)}.
\end{equation}
For instance, this places the freezing transition (estimated by
setting $\eta=0.5$) at $\Gamma\approx122$ which is within 40\% of the
known value $\Gamma\approx180$. Stroud and Ashcroft have examined the
accuracy of this approximation in some detail and found that it is
good provided $\Gamma>10$ or so.\cite{SA}

To summarise using the classification in the preceeding section, the
free energy in Eq.~\eqref{ftoteq} is valid for colloidal systems in
regions II and III provided the ratio $Z\lB/\sigma$ is not too large.

\section{Phase behaviour}
The most remarkable thing about the free energy in Eq.~\eqref{ftoteq}
is that it can have a region of negative osmotic compressibility,
indicating that a phase instability is present.  In a phase diagram,
this appears as a closed-loop miscibility gap.  Before discussing this
in detail, it is worthwhile noting that \emph{all three} electrostatic
pieces act to destabilise the system, just as Langmuir surmised. This
includes the macroion-macroion interaction term which is usually
supposed responsible for the system's overall thermodynamic
stability. But it too is destabilising because of the background
subtraction counterterm. A closer examination (see below) shows that
it is the ideal translational entropy of the counterions, and to a
lesser extent the hard core exclusion term, that are responsible for
the system's stability. A miscibility gap opens up where these
stabilising mechanisms fail.

\subsection{Phase instability or miscibility gap}
\begin{figure}
\begin{center}
\vspace{15pt}
\includegraphics{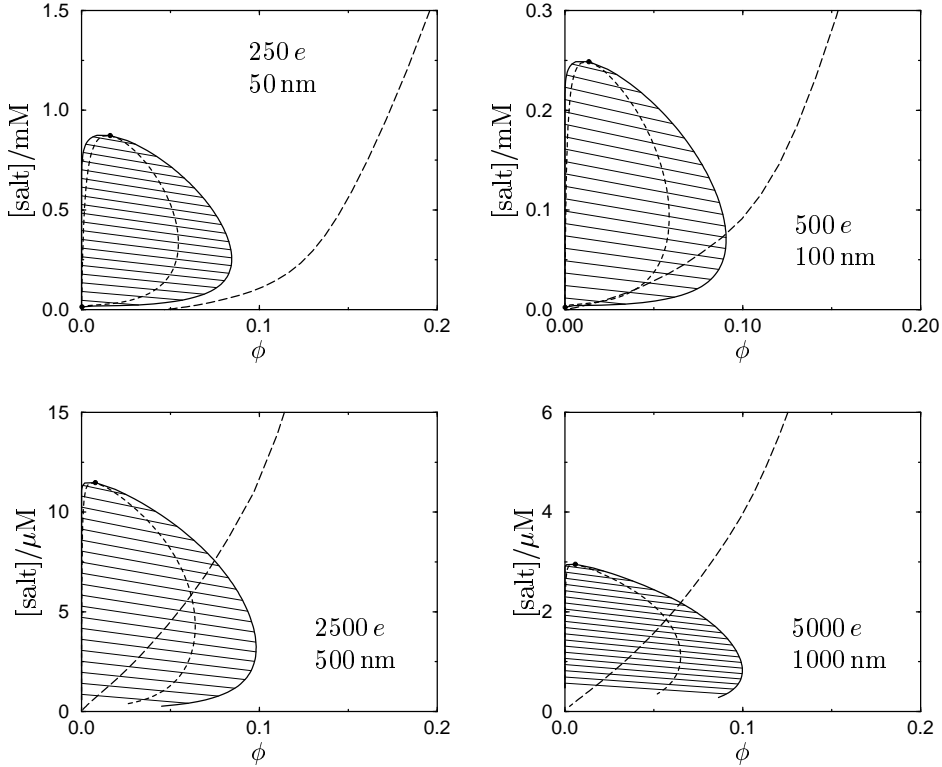}
\end{center}
\caption[?]{Phase diagrams showing closed loop miscibility gaps for
several colloid particle sizes, at charges roughly corresponding to
$\Zmax\approx4\sigma/\lB$.  The solid lines are binodals and tie
lines.  The short-dashed lines are spinodal lines.  The long-dashed
lines show where the variational volume fraction is 50\%, and give an
indication of the location of the freezing transition.  See
Table~\ref{tab:ucsp} for details of the upper critical solution
point.\label{fig1}}
\end{figure}

\begin{figure}
\begin{center}
\vspace{15pt}
\includegraphics{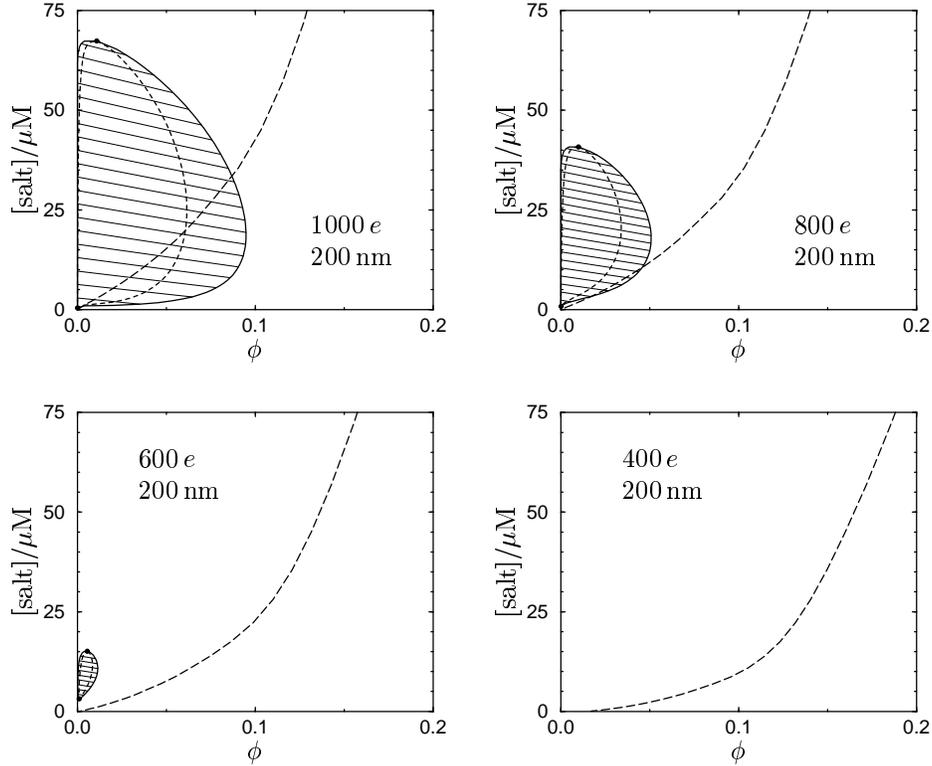}
\end{center}
\caption[?]{As the charge on the colloid particles is reduced, the
miscibility gap shrinks and finally disappears for $Z<\Zmin\approx565$
in this case. At $Z=\Zmin$ the upper and lower critical solution
points meet at a double critical point or hypercritical
point.\label{fig2}}
\end{figure}

Within the miscibility gap, coexistence compositions are calculated
by standard methods corresponding to conditions of equality of
chemical potential and osmotic pressure between coexisting phases. In
deriving expressions for the chemical potentials and osmotic pressure,
care has to be taken to account fully for the state point dependence
of all parameters such as $\kappa$ in Eqs.~\eqref{fsmseq}, etc, and
$\lambda$ and $\eta$ in Eq.~\eqref{fmmeq}, and also the fact that the
free energy should be evaluated at a variational minimum with respect
to $\eta$. Phase diagrams are shown in the $(n_m, n_s)$ plane.
Typical results are shown in Figs.~\ref{fig1} and \ref{fig2}.

The interpretation of these diagrams is standard.  Outside the
miscibility gap, a colloidal suspension is predicted to remain stable
and homogeneous.  Within the miscibility gap, an initially homogeneous
suspension is predicted to phase separate into colloid rich and
colloid depleted regions along the tie lines indicated in these
diagrams.  The relative amounts of the two phases will be determined
by the lever rule.  Typically, though, the colloid volume fraction in
the depleted phase is vanishingly small and this phase is essentially
pure brine.  Also, if the mean colloid volume fraction is nearer the
right hand binodal, the second phase will appear as droplets of pure
brine within the colloid-rich phase.  This, it is argued, is a natural
explanation for the void structures described in the
introduction.\cite{voids,paradox}

Within the miscibility gap one finds a spinodal region, demarcated by
the short-dashed lines in Figs.~\ref{fig1} and \ref{fig2}. In this
region, the osmotic compressibility is negative, indicating spinodal
decomposition will occur.\cite{spinbad}  Also shown in
Figs.~\ref{fig1} and \ref{fig2} are long-dashed lines where the
variational volume fraction $\eta=0.5$.  This is an indication of the
location of the freezing transition to an ordered colloidal crystal
phase.  The width of this transition is determined essentially by the
compressibility of the supporting electrolyte,\cite{Stishov} and is
expected to be narrow in general, but widen considerably as the
spinodal line in the miscibility gap is approached.  This is indeed
found by van Roij, Dijkstra and Hansen\cite{vRDH} whose calculations
are superior in this respect to the present theory, since they
explicitly include the possibility of an ordered phase.  The proximity
of the freezing transition is not entirely a coincidence, since the
miscibility gap occurs close to the point where the system starts to
behave as a strongly coupled OCP (region II) where the plasma coupling
constant $\Gamma\sim Z$.  Experimentally, it has also been noted that
anomalous behaviour in the static structure factor starts close to the
point where the ordered phase appears at low ionic strengths.\cite{USAXS}

In general terms, the miscibility gap appears at low ionic strengths,
low volume fractions and high macroion charges. Fig.~\ref{fig1} shows
what happens to the miscibility gap for various macroion diameters, at
charges roughly corresponding to $4\sigma/\lB$ which is the estimate
of the maximum charge at which the linearised DH approximation for
the macroion polarisation energy starts to break down severely.  As
the macroion size increases, the miscibility gap moves to lower salt
concentrations, but remains approximately in the same place with
regard to macroion volume fraction.

The miscibility gap is bounded above and below by critical solution
points. Note that in the present theory, a strictly salt-free system
has no phase instability.  As the charge on the macroions is reduced,
the miscibility gap reduces in size and finally vanishes once the
charge falls below some minimum value $\Zmin$, illustrated in
Fig.~\ref{fig2} for one particular macroion size. At the point where
the miscibility gap vanishes, the upper and lower critical solution
points coincide. This point is identified as a double critical point
or hypercritical point.\cite{clmg}

\begin{figure}
\begin{center}
\vspace{15pt}
\includegraphics{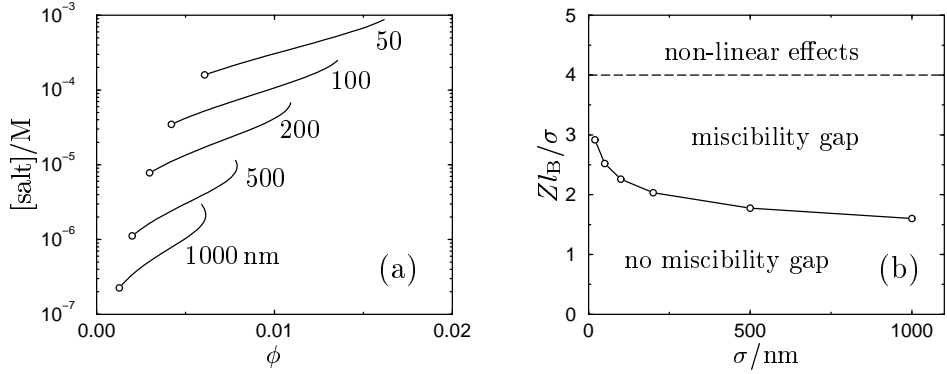}
\end{center}
\caption[?]{(a) Fate of the upper critical solution point as $Z$ varies
between $\Zmax\approx4\sigma/\lB$, and $\Zmin$ where it meets the lower
critical solution point (open circles). Data for various particle
diameters are shown. (b) State diagram showing where a miscibility gap
is found. The open circles indicate the points where $Z=\Zmin$ for
various $\sigma$, where the miscibility gap just disappears. Essentially
this is a line indicating where double critical points occur in the
phase diagram (see Table~\ref{tab:dcsp}). The dashed line shows where
$Z=4\sigma/\lB$, which is one possible criterion for acceptability of
the \DH\ approximation for the polarisation energy.\label{fig3}}
\end{figure}

The upper critical solution point can be taken as approximately
representative of the position of the miscibility gap as a whole, and
Fig.~\ref{fig3}(a) shows what happens to it as $Z$ varies between $\Zmax
\approx 4\sigma/\lB$ and $\Zmin$, for various particle diameters. As
$Z$ is reduced, in general the upper critical solution point moves to
lower volume fractions and ionic strengths, before vanishing at the
double critical point.

Fig.~\ref{fig3}(b) shows the region in parameter space where the
phenomenon of a closed-loop miscibility gap is found. The solid line
with circles in this plot indicates where the miscibility gap just
disappears, where a double critical point arises in the phase
diagram. The behaviour is plotted in the space $(\sigma,Z\lB/\sigma)$
which serves to bring out an important point. At low macroion size,
the value of $\Zmin$ required to observe the miscibility gap
approaches $\Zmax$, where non-linear effects start to become
important. This may explain why anomalous effects are only observed
for larger colloid particles where a sufficient gap opens up between
$\Zmin$ and $\Zmax$ for the phenomena to be experimentally accessible.

Some representative numerical data on the upper critical solution
points are collected in Table~\ref{tab:ucsp}, and the double critical
points in Table~\ref{tab:dcsp}.  These tables show the small ion Debye
screening length $\kappa^{-1}$ in relation to other lengths in the
problem, at the critical point in question.  The first point to note
is that the upper critical solution point always lies approximately at
$\kappa a\sim1$.  This fact has been noticed
before\cite{USAXS,SSH2,SSH3} and interpreted as evidence in support of
the SI theory (see appendix)---for example one might compare the
measured $\kappa X_{\mathrm{min}}$ of Ref.~\onlinecite{SSH3} with
$\kappa\ro$ in Table~\ref{tab:ucsp}, although the two systems are not
the same.  Here $\kappa a\sim1$ emerges naturally and is seen to
reflect the particular importance of the finite macroion size
correction factor, $f(x)$ in the second of Eqs.~\eqref{fsmseq}, around
$x=\kappa a\sim1$ (see also Fig.~\ref{potfig}).  The second point to
note is that, to $O(Z)$, the miscibility gap occurs at $\kappa\ro\sim
Zn_m/n_s\sim1$.  This places the miscibility gap at the apex of region
II where it meets with regions III-a and III-b in the state diagram in
Fig.~\ref{regfig}.


\vspace{24pt}

\begin{table}
\begin{tabular}{rrcccdddd}
&&&\multicolumn{2}{c}{Upper critical solution state-point}\\
$\sigma/\nm$&$Z$&$Z\lB/\sigma$&$\phi$&$\salt/\M$&%
  $\kappa^{-1}/\nm$&$\kappa a$&$\kappa\ro$&$2n_s/Zn_m$\\
\tableline
  50&  250& 3.60& $1.62\times10^{-2}$&%
   $8.74\times10^{-4}$&   9.96& 2.51&  9.92&17.0\\
 100&  500& 3.60& $1.36\times10^{-2}$&%
   $2.49\times10^{-4}$&  18.8 & 2.66& 11.1 &23.1\\
 200& 1000& 3.60& $1.09\times10^{-2}$&%
   $6.74\times10^{-5}$&  36.3 & 2.75& 12.4 &31.1\\
 200&  800& 2.88& $9.93\times10^{-3}$&%
   $4.08\times10^{-5}$&  46.6 & 2.15&  9.99&25.9\\
 200&  600& 2.16& $5.51\times10^{-3}$&%
   $1.51\times10^{-5}$&  76.3 & 1.31&  7.42&23.0\\
 500& 2500& 3.60& $7.82\times10^{-3}$&%
   $1.15\times10^{-5}$&  88.5 & 2.83& 14.2 & 46.3\\
1000& 5000& 3.60& $5.91\times10^{-3}$&%
   $2.95\times10^{-6}$& 175.  & 2.86& 15.8 & 63.0\\
\end{tabular}
\vspace{24pt}
\caption[?]{Location of upper critical solution point for various macroion
charges and diameters. Also shown is the Debye screening length and its
relation to the macroion radius $a$ and mean spacing between macroions
$\ro$. The last column gives the ratio of contributions to $\kappa$ from
the salt ions ($2n_s$) and the macroion counterions ($Zn_m$), at the
state point in question.\label{tab:ucsp}}
\end{table}

\vspace{24pt}

\begin{table}
\begin{tabular}{rrcccdddd}
&&&\multicolumn{2}{c}{Double critical solution state-point}\\
$\sigma/\nm$&$\Zmin$&$\Zmin\lB/\sigma$&$\phi$&$\salt/\M$&%
  $\kappa^{-1}/\nm$&$\kappa a$&$\kappa\ro$&$2n_s/Zn_m$\\
\tableline
  50&  175& 2.52& $6.08\times10^{-3}$&%
   $1.59\times10^{-4}$&  23.0& 1.09& 5.95& 11.8\\
 100&  314& 2.26& $4.21\times10^{-3}$&%
   $3.48\times10^{-5}$&  49.9& 1.00& 6.21& 16.6\\
 200&  565& 2.03& $2.99\times10^{-3}$&%
   $7.78\times10^{-6}$& 106. & 0.94& 6.53& 23.3\\
 500& 1233& 1.78& $2.01\times10^{-3}$&%
   $1.12\times10^{-6}$& 282. & 0.89& 7.02& 35.6\\
1000& 2225& 1.60& $1.28\times10^{-3}$&%
   $2.25\times10^{-7}$& 633. & 0.79& 7.27& 49.6
\end{tabular}
\vspace{24pt}
\caption[?]{Location of double critical solution points for various
macroion diameters, which occur when the upper and lower critical
solution points coincide and the miscibility gap just appears, at
$Z=\Zmin$.\label{tab:dcsp}}
\end{table}

These results are in broad agreement with recent detailed experiments
of Yoshida, Yamanaka and co-workers.\cite{bi1,bi2} For instance Fig.~2
in Ref.~\onlinecite{bi2} indicates that for $\sigma\approx120\,\nm$,
$Z\approx200$ particles (sample KE-P10W), a biphasic region appears
around $\phi\approx1$--5\% and $n_s\approx10$--$20\,\mu\M$.  For these
values of $\sigma$ and $Z$, the present theory does not predict a
miscibility gap, although it does do so at somewhat higher $Z$
(Table~\ref{tab:dcsp}).  However, one would expect a region of greatly
lowered osmotic compressibility to persist in that area of the phase
diagram for smaller values of $Z$.  If the freezing transition comes
close to such a region, one would expect it to broaden
considerably.\cite{Stishov} This is seen for instance in the
calculations of RDH who include the possibility of an ordered
phase. The measured biphasic region lies at
$Zn_m\approx(0.2$--$1)\times10^{-5}\,\nm^{-3}$ and
$n_s\approx(0.5$--$1)\times10^{-5}\,\nm^{-3}$, both of which are close
to the apex of region II in Fig.~\ref{regfig}, located at
$Z^{-2}\lB^{-3}\approx7\times10^{-5}\,\nm^{-3}$.  To $O(Z)$, this is
the same place that the miscibility gap appears in the present theory.
Whilst far from conclusive, one can argue that this is evidence the
theories are on the right track.  Similarly one can calculate
$\kappa^{-1}\approx80\,\nm$ and thus $\kappa a\approx0.7$, in
agreement with the above general observations. Some of the detailed
trends reported by Yoshida \etal\ are not seen in the present theory,
perhaps because detailed account is not made of the ordered
phase. Similarly, there are remarkable kinetic phenomena which are not
covered in this equilibrium theory, and which I will touch on briefly
again below.

\subsection{Relative importance of free energy contributions}
We can determine the relative importance of the various contributions
to the free energy in Eq.~\eqref{ftoteq} simply by recalculating the
phase instability regions in the presence or absence of one or more of
these terms.  In this way the following sequence of importance is
discovered:
\begin{equation}
F_{\text{id}}, F_{\text{ms}} > F_{\text{mm}} > F_{\text{hc}} 
\gg F_{\text{ss}}.
\end{equation}
The driving force for phase separation is mainly due to the
polarisation term, the second of Eq.~\eqref{fsmseq}, and to a lesser
extent the macroion interaction term Eq.~\eqref{fmmeq}, whereas it is
the ideal translational entropy term Eq.~\eqref{fideq} that is the
main driving force for stability.  When the translational entropy is
insufficient to balance the electrostatic free energy, a miscibility
gap appears.  The relative importance of the macroion-small ion
interaction contribution has been confirmed very recently by direct
numerical simulation.\cite{LoLi}

If any of the polarisation term, the macroion interaction term, or the
background subtraction counterterm in the latter, is omitted, the
phase instability moves to much higher $Z$ and lower salt
concentration.  On the other hand, the presence or absence of the
small ion electrostatic free energy, the first of Eqs.~\eqref{fsmseq}
has essentially no effect on the location of the phase transtion.
This is because it has a similar state point dependence to the
polarisation term but is diminished in magnitude relative to this by a
factor $Z$.

The details of the miscibility gap also rely on the finite size
correction factor appearing in the second of Eq.~\eqref{fsmseq}.  If
$f(x)$ is set to unity in this, the phase instability may not be
bounded from above in $n_m$.  In this respect, the present theory
resembles the finite-size corrected DH theory of simple
electrolytes.\cite{Fisher} Similarly we can check the influence of
the finite size corrected macroion charge in the macroion interaction
term (a factor $e^{2\kappa a}/(1+\kappa a)^2$ in Eq.~\eqref{fmmeq}).
The presence or absence of this factor is found to have only a weak
effect on the location of the phase transition.  Other finite size
corrections might be applied for instance to the screening length
expression itself, to take account of the increased concentration of
small ions due to the volume excluded by the macroions.  These effects
have not been seriously explored since $\phi$ is small at the point
where the phase instability appears.  On the other hand, since the
small ion entropy is so significant, the factors $1/(1-\phi)$
appearing in Eq.~\eqref{fhceq} do play a \role\ in suppressing the
phase instability at high $\phi$.

The \role\ of the ideal translational entropy of the counterions,
in stabilising the suspension at higher salt concentrations, can
be seen in an elementary way by noting that at high $n_s$
the small ion entropy term in Eq.~\eqref{fideq} becomes
\begin{equation}
(Zn_m+n_s)\log(Zn_m+n_s)\approx
{\mathrm{consts.}}+\frac{Z^2n_m^2}{n_s}.
\end{equation}
It is seen that this contributes an effective positive term $Z^2/n_s$
to the effective second virial coefficient between the macroions.
Since $Z\gg1$ this represents a strong stabilising influence.  Note
also that if $n_s\gg Zn_m$, the screening length is dominated by $n_s$
and is not strongly dependent on $n_m$.  This means that the macroion
polarisation term becomes approximately linearly proportional to
$n_m$, and acts simply to shift the macroion chemical potential,
rather than destabilise the suspension.

The conclusion is that the appearance of the miscibility gap is driven
by the electrostatic free energy but opposed principally by the
translational entropy of the counterions ions.  Remarkably, it is
precisely these two effects which were identified by Langmuir in his
1938 analysis.  A similar conclusion has also been reached for charged
plates recently.\cite{SSS} The exact location and width of the
miscibility gap depends on a subtle combination of finite size effects
though, and in contrast to Langmuir's conclusion, the general
stability of a colloidal suspension does not depend on anything more
esoteric than the domination of small ion entropy in the free energy.

\begin{figure}
\begin{center}
\vspace{15pt}
\includegraphics{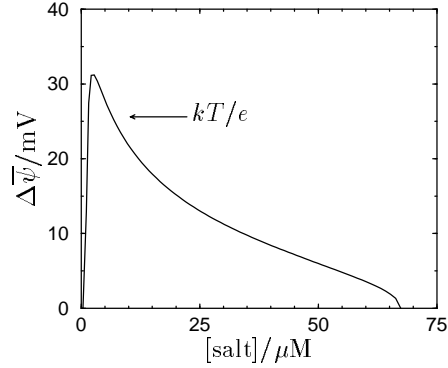}
\end{center}
\caption[?]{Donnan potential between coexisting phases for $Z=1000$
and $d=200\,\nm$ (see Fig.~\protect\ref{fig2} for actual phase
diagram), plotted as a function of the salt concentration in the
macroion-poor phase, which is essentially pure brine.  The sign is
such that the macroion-rich phase is at a higher potential than the
macroion-poor phase.\label{donfig}}
\end{figure}

\subsection{Donnan potential difference}
The phase behaviour has been calculated using $n_m$ and $n_s$ as
dependent variables.  An alternative is to use $n_m$, $\np$ and $\nn$
as variables and replace the electroneutrality condition by a Donnan
potential term,
\begin{equation}
\frac{F_{\text{Donnan}}}{V\kT}=\frac{e\potD}{\kT}(Zn_m+\np-\nn),
\end{equation}
addded to the free energy density.  The Donnan potential or mean
electrostatic potential in each phase, $\potD$, is chosen to ensure
electroneutrality in any bulk phase.\cite{DH} One could view it
as a Lagrange multiplier for the electroneutrality condition, although
it does have a physical significance.  Since this potential will in
general be different in coexisting phases, it means that there is in
general a \emph{Donnan potential difference}, $\Delta\potD$, between
coexisting phases.  This is like a membrane equilibrium, where the
interface between the two phases plays the \role\ of the 
membrane.\cite{SSH1}  Clearly, much insight into the nature of the phase
transition can be gained by examining the structure of this
interfacial region, and a knowledge of $\Delta\potD$ throws some light
on this.

It is fairly easy to derive an expression for $\Delta\potD$. First one
can show that
\begin{equation}
\mu_\smallpm=\pm e\potD+\kT\log (f\!n_\smallpm)\label{siaeq}
\end{equation}
where the activity coefficient $f$ is the \emph{same} for both species
since the excess free energy only depends on the combination
$(\np+\nn)$.  From the constancy of $\mu_\smallplus-\mu_\smallminus$,
the Donnan potential difference between a pair of coexisting phases is
\begin{equation}
\Delta\potD=\potD^{\mathrm{(ii)}}-\potD^{\mathrm{(i)}}
=\frac{1}{2}\frac{\kT}{e}
\biggl[\log\biggl(1+\frac{Zn_m}{n_s}
\biggr)\biggr]_{\mathrm{(i)}}^{\mathrm{(ii)}}.
\end{equation}
This allows easy determination of $\Delta\potD$ given the compositions
of a pair of coexisting phases, and a typical result is shown in
Fig.~\ref{donfig}.  Note that the macroion-rich phase is at a higher
potential than the macroion-poor phase, and given the previous
observation that $Zn_m\sim n_s$, the potential difference between
the phases is of order $\kT/e\approx25\,\mathrm{mV}$.

\begin{figure}
\begin{center}
\vspace{15pt}
\includegraphics{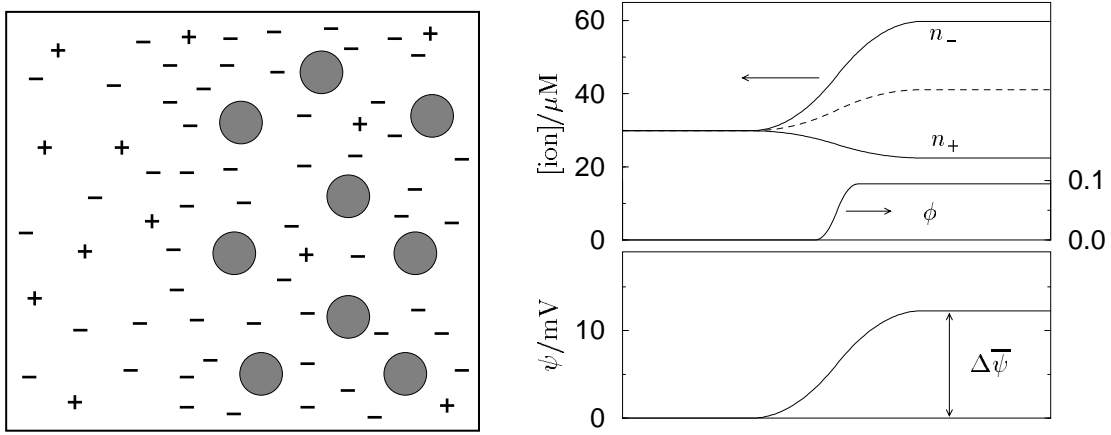}
\end{center}
\caption[?]{Postulated structure of the interface between two
coexisting phases in a miscibility gap.  The macroion-poor phase (left
hand side) is essentially pure brine.  The macroion-rich phase has a
large excess of counterions to match the electrostatic charge of the
macroions.  These spill across the interface and set up a macroscopic
double layer raising the mean potential in the macroion-rich phase and
consequently depleting it of coions.  The dashed line is the ionic
strength, $(\np+\nn)/2$.  The known coexistence compositions have
been smoothly connected across the interface here.\label{interfig}}
\end{figure}

\section{The nature of the interfacial region}
As mentioned above, much insight into the nature of the above phase
transition can be gained by studying the interface between coexisting
phases.  Such a study would require the generalisation of the present
theory to inhomogeneous states.  This has not yet been attempted, but
is clearly possible, perhaps using methods described by Rowlinson and
Widom,\cite{RowWid} density functional theory,\cite{BH,GED} and the
generalised DH theory of Lee and Fisher for the
polarisation free energy in a salt gradient.\cite{LF}  Nevertheless, there is
enough information in the present calculations to make an informed
guess as to the structure of the interfacial region in this charged
system.

Fig.~\ref{interfig} shows a postulated structure of the interface
between two co-existing phases in the middle of the miscibility gap
for $\sigma=200\,\nm$ and $Z=1000$ (see Fig.~\ref{fig2} for actual
phase diagram).\cite{NN}  In drawing this structure, one is guided by
the calculated Donnan potential difference between the two phases.
This potential difference can only be explained by a macroscopic
electric field across the interface.  The sign of the potential
difference indicates that a kind of macroscopic super-double layer
forms because of an excess of counterions spilling out into the
macroion-dilute phase.  In fact, this is entirely to be expected from
consideration of the translational entropy terms discussed above.  The
co-ions are depleted from the macroion rich phase by the well known
Donnan common ion effect (from Eq.~\eqref{siaeq} we see that the
product of the small ion activities $(f\!\np)(f\!\nn)$ is common to
the two coexisting phases).  Since the co-ions can be taken as
representing the salt concentration, this also corresponds to the salt
partitioning seen in the full phase diagrams.  Although the salt
partitions to some extent out of the macroion rich phase, it is easy
to check that the ionic strength, $(\np+\nn)/2$, is increased in this
phase.

Before leaving this subject, one should mention that it is entirely
possible for oscillatory density profiles to develop across the
interface, instead of the smoothly interpolated densities suggested by
Fig.~\ref{interfig}.  This might be expected in the neighbourhood of
the critical point, corresponding to the possible appearance of charge
density wave phases mentioned below.\cite{NNP}

\section{Many-body attractions?}
The theory thus presented is self contained, yet it seems to the
present author that the story, if it stops here, is not quite
complete.  The phase instability appears despite the fact that the
macroion pair interactions remain repulsive, as discussed at length by
RDH.\cite{vRH,vRDH} The instability is driven by volume terms which
don't feature in the pair potentials.  Simple arguments below though
suggest that the volume terms should have an alternative
interpretation in terms of an \emph{attractive many-body interaction}.
The interpretation of the void phenomena by many-body attractions has
been championed by Schmitz,\cite{Schmitz1} who considers the
interactions of a macroion with a collection of other macroions at set
positions in a \PB\ theory.  The arguments below are partly motivated
by these calculations.

Of course, it is well known that an attractive component can appear in
the pair interactions due to correlation effects not captured in the
present mean field theories.  In the simplest picture, the attraction
arises from correlated fluctuations of the ion clouds around the
macroions, and is the classical analogue of the London-van der Waals
forces.  The effect can be very significant for multivalent
electrolytes,\cite{StRo,Patey} and in certain circumstances can be
sufficiently strong to drive phase separation.\cite{LiLo} For
colloidal systems in the present regime though, this mechanism seems
ruled out.\cite{LiLo,DLVOexpt}

Beyond the pair level, a hint of the existence of many-body
attractions is provided by experiments which seem to show an
attractive interaction between pairs of charged colloid particles in
the presence of wall(s).\cite{LG,twocollwall} Theoretically, the
situation in these geometries is not clear cut.  Non-linearities in
\PB\ theory\cite{BoSh,QB} seemed to provide an explanation of these
observations until challenged by subsequent work\cite{JCN} and density
functional theory\cite{TATKB} which argue for the absence of
attractions.  However other theoretical work,\cite{GH} and recent
simulations,\cite{AAL} seem to show that attractions might appear
beyond mean field theory.

Let us discuss two arguments for the existence of many-body
attractions, even in the mean field theory.  The first argument is
essentially due to Smalley.\cite{Smalley} Start with a homogeneous
distribution of macroions such that the mean composition lies within
the two phase region identified above.  Next move the macroions
collectively so that we have a void region coexisting with a
homogeneous distribution of macroions at an enhanced density and the
appropriate ordered crystal or strongly correlated fluid structure.
If the compositions and relative volumes of the two regions are chosen
correctly, the final state will correspond to a pair of coexisting
state points in the phase diagram.  The free energy must decrease in
such a process, therefore there \emph{must} be a many-body
thermodynamic force tending to accumulate macroions in the manner
suggested.  This argument is rather general, and indicates a $N$-body
attraction appears for $2<N<10^{11}$ or so (see below for a more
refined estimate).

\begin{figure}
\begin{center}
\vspace{15pt}
\includegraphics{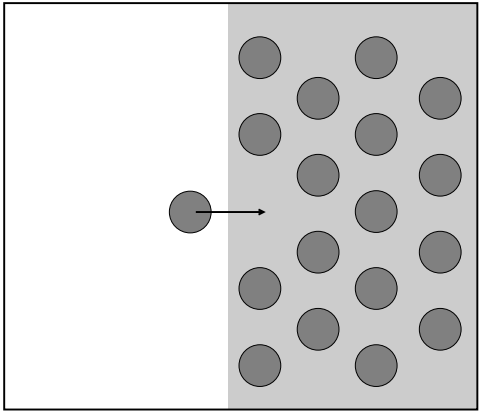}
\end{center}
\caption[?]{A general argument in the text indicates that, in the
interfacial region, there should be a net attraction of an isolated
macroion to a correlation hole in the macroion-rich
phase.\label{forcefig}}
\end{figure}

The second argument concerns the forces acting on macroions in the
vicinity of the interfacial region.  Note that this region does
resemble the experimental situation mentioned
above,\cite{LG,twocollwall} given an excess of counterions in vicinity
of the interfacial region that extends some way into bulk of the
macroion-poor phase.  Thus an attraction between pairs of macroions
adjacent to the macroion-rich phase might be expected.  However, in
the interfacial region, there ought to be a net attraction between an
\emph{isolated} macroion and the macroion-rich phase. To see this,
imagine introducing a tracer macroion into the interfacial region.  In
equilibrium, the probability distribution, $P(z)$, for the position of
the tracer particle across the interface should be proportional to the
macroion number density profile, $n_m(z)$, otherwise the particle
isn't doing its job as a tracer.  Now consider $P(z)$ as the
stationary solution of some Smoluchowski equation
\begin{equation}
\frac{\partial P}{\partial t}=\frac{\partial}{\partial z}
\biggl(\frac{1}{\xi}\biggl(\kT\frac{\partial P}{\partial z}+
P\frac{\partial U}{\partial z}\biggr)\biggr)
\end{equation}
in which $\xi$ is the local friction coefficient for the tracer
particle, and $-\partial U/\partial z$ is the force acting on the
tracer macroion due to the interfacial profile structure.
Inserting $P(z)\propto n_m(z)$ and requiring stationarity shows that
this force is simply $(\kT/n_m)\partial n_m/\partial z$.  Since
the macroion density increases through the interface, this means that
there must be \emph{a net attraction between the tracer particle and the
macroion-rich phase}.

What is the origin of this force? Clearly we can decompose it into a
number of effects. The electric field (potential gradient
$\partial\pot/\partial z$) acts to \emph{repel} the tracer macroion
from the macroion-rich phase, since the tracer is positively
charged. However there is also a gradient in the ionic strength
(Fig.~\ref{interfig}), leading to a force attracting the tracer
\emph{towards} the macroion-rich phase, due to the polarisation energy
discussed earlier (Fig.~\ref{potfig}). In fact, in this case we can
show that this is not quite sufficient to overcome the electrostatic
repulsion, essentially because the miscibility gap does not appear at
these parameter values if only the polarisation term is included in
the free energy.\cite{attrnote}

The final contribution which tips the balance in favour of an
attraction towards the macroion-rich phase derives from the
macroion-macroion correlations represented by the free energy
Eq.~\eqref{fmmeq} in the bulk.  Note that the background subtraction
counterterm in this corresponds to the fact that there is no net
interaction between the tracer and macroions deep in the bulk of the
macroion-rich phase.  The only force acting over such large distances,
the Coloumb force, is exactly cancelled by the net charge of the small
ions since the bulk is electrically neutral (which is another way to
see that $Z$ rather $\Zeff$ should appear in the background
counterterm).  The macroion-macroion contribution to the force
therefore corresponds to a pure correlation effect: the electrostatic
repulsion between the tracer and neighbouring macroions causes them to
move away leaving an excess of counterions in the vicinity of the
tracer. It is the attraction of the tracer to this `correlation hole',
the centre of which is displaced towards the macroion-rich phase in
the interfacial region, that appears to be the final piece in the
jigsaw.  This is illustrated in Fig.~\ref{forcefig}, but clearly the
lack of a proper theory for the inhomogeneous state is most keenly
felt in formulating this notion more precisely.

In the theory for void formation presented above, or the RDH theory,
these many-body effects are apparently not
present.  In fact a kind of many-body interaction has crept in by
sleight-of-hand via the electroneutrality condition.  This condition
means that it is impossible to vary the macroion number density in the
bulk without a concomitant variation in the small ion densities.  The
consequent effect, via changes to $\kappa$ for instance, feeds back
through otherwise innocuous terms in the free energy to influence the
thermodynamic properties of the system in a non-trivial manner.  As
the discussion on the interfacial region shows though, the
electroneutrality condition corresponds to the appearance of a Donnan
potential difference between the phases, hence an electrostatic
potential gradient and a macroscopic double layer.  Therefore the
rather innocent-looking electroneutrality condition conceals some
rather important physical effects.  In fact, Sogami and Ise already
identified the crucial importance of this condition---see
appendix.\cite{SI}

Whilst in an ordinary liquid, many-body interactions are generally
insignificant, the same conclusion is not necesarily true in ionic
systems.  Although pair interactions may be screened, the long range
nature of the Coulomb law is never far from the surface, and reappears
on whatever length scale is necessary as soon as inhomogeneous charge
distributions arise.

The co-operative nature of many-body attractions may shed some light
on one of the mysteries of the void phenomenon mentioned in passing
above, namely the length of time it appears to take for voids to form.
It is easy to estimate times for macroions to diffuse distances of
order the interparticle spacing.  These times ($\ro^2/D$) are of the
order of 10--100 seconds or less, even for the largest ($1\,\um$)
particles, whereas the void formation time is measured in 10--100
hours.\cite{voids} This certainly appears to present a difficulty for
theories based on attractive minima in pair potentials, and arguments
which rely on a time scale separation between the small ion and
macroion dynamics,\cite{SI,LBT} but can be explained rather easily in the
present theory.  For example, in the spinodal instability region, one
might expect that only density fluctuations of sufficiently long
wavelengths $\lambda$ to incorporate sufficiently many macroions will
be unstable and grow.  The growth rate decreases as $\lambda^{-2}$,
and the difference in the above time scales (a factor of $10^3$ or so)
can be explained by supposing that the minimum fluctuation size is of
order $(10^3)^{1/2}\sim30$ or so particles in linear dimensions.  In
the above terms, this implies $N$-body attractions start to be
significant around $N\sim30^3\sim10^4$ particles.

At present though, no immediate explanation seems forthcoming for the
elaborate time sequences of phase transformations that have sometimes
been observed.\cite{YYKIH,bi1} Uncharged colloid-polymer mixtures with
vapor-liquid and freezing transitions in their phase diagram have
recently been demonstrated to have a rich kinetic
behaviour.\cite{WCKP} In the present case where the vapour-liquid
phase transition is driven at least in part by many-body attractions,
even more varied novel kinetic pathways seem possible.

\section{General discussion}
We have seen how a carefully justified \DH\ (DH) theory can be constructed
for the free energy of a charge stabilised colloidal suspension,
treated as a highly asymmetric electrolyte or plasma.  Linearisation
and mean-field approximations are justified and applied to the
interactions of small ions with small ions, and of small ions with
macroions. These approximations cannot be supported for the
macroion-macroion interactions, which are treated separately by a
well-established variational method.

The most remarkable prediction that emerges from the analysis is the
appearance of a phase instability in the form of a closed-loop
miscibility gap, at low ionic strengths, low volume fractions, and
high macroion charges.  This gap appears to be the analogue, for the
highly asymmetric electrolyte, of the vapour-liquid coexistence region
in the RPM.  The present theory confirms the recent calculations of
van Roij, Dijkstra and Hansen (RDH) who use a different approximation
method.\cite{vRH,vRDH} Void formation and other anomalies in highly
charged colloidal suspensions at low ionic strengths would appear to
have a natural explanation in terms of this phase instability.

The phase instability is driven by the electrostatic correlation free
energy, and is principally opposed by the counterion translational
entropy acting in concert with the condition of bulk phase
electroneutrality.  The precise location and extent of the instability
depends on more subtle details, such as finite macroion size
correction factors in the various terms in the free energy.

The above analysis has not exhausted the list of possible phases that
might form in charge stabilised colloidal suspensions.  As well as
ordered colloidal phases (colloidal crystals) that have been included
by RDH there is a rather general argument based on a Landau theory
that \emph{charge-density-wave} phases might be expected in the
vicinity of critical points in charged systems,\cite{NNP} in other
words in the vicinity of the upper and lower critical solution points
in the present theory.  This possibility arises because such modulated
phases can break the electroneutrality condition that applies in bulk
phase separation, and allow the density of small ions to be more
uniformly spread than the macroion density.  To decide whether such a
phase is stable or not requires that the present theory be elaborated
to treat such inhomogeneous states though.

The extension to treat inhomogeneous states is also a critical step in
constructing a theory for the interfacial region, from which many
insights might be gained into the nature of the phase transition and
the role of many-body attractions, which are concealed to some extent
in the homogeneous theory by the use of the electroneutrality
condition.  

The theory can be extended in a number of other directions too.  One
extension is to improve the DH approximation for the polarisation free
energy taking into the saturation effect indicated in
Fig.~\ref{grootfig}, and avoid the charge that the interesting effects
in the present theory lie at the margin of its validity (see appendix
C).  Another extension is to see what happens if the present
fixed-charge macroions are replaced with fixed-surface-potential
macroions, or replaced with a more physical charge-regulating model
for the surface,\cite{chreg} thereby addressing Langmuir's point (C)
in the introduction.

One of the most interesting developments is to replace the supporting
simple electrolyte by a more complex fluid, such as a polyelectrolyte
solution. Let us suppose that the polyelectrolyte chains have an
opposite charge from the macroions. Whilst there are effects on all
components in the electrostatic free energy, perhaps the single most
important effect might be the reduction in importance of the
translational entropy of the counterions.\cite{PBW,Linse} This can be
captured by replacing $(Zn_m+n_s) \log(Zn_m+n_s)$ of Eq.~\eqref{fideq}
by $(n_p/N_p) \log(n_p/N_p) + (Zn_m+n_s-fn_p) \log(Zn_m+n_s-fn_p)$,
where $n_p$ is the additional polyelectrolyte segment density, $N_p$ is
the degree of polymerisation, and $f$ is the charge per segment.
Preliminary investigations show that the effect of the polyelectrolyte
in this model is to strongly amplify the phase instability, particularly
around the charge stoichiometry plane $Zn_m=fn_p$, in accordance with
expectations. This simple picture might lead to new insights into the
phase behaviour (coacervation) of mixtures of oppositely charged
polyelectrolytes and colloids\cite{Peff} or proteins.\cite{polprot}

A final extension might be to re-introduce van der Waals attractions
between macroions.  This should result in the re-appearance at high
salt concentrations of a second phase instability, corresponding to
the salting out phenomenon presently captured by the DLVO theory.
Indeed such a theory has been constructed by Victor and
Hansen,\cite{VH} and Grimson \etal,\cite{GMS} for related models.
In this way, one might hope for a truly general theory that
encompasses \emph{all} aspects of the stability of lyophobic colloids.

\section{Acknowledgments}
For useful discussions and correspondence, I thank M. E. Cates, R.
Evans, R. D. Groot, J.-P. Hansen, M. Lal, Y. Levin, K. S. Schmitz,
R. P. Sear, M. V. Smalley, I. D. Robb, R. van Roij.  I particularly
thank D. G. Hall for initiating me in the mysteries of charged
colloidal suspensions, some years ago now, and drawing my attention to
Langmuir's work.

\appendix
\section{Comparison with the Sogami-Ise theory}
By making a distinction between the `Helmholtz pair potential' and the
`Gibbs pair potential', Sogami and Ise (SI) introduce an ingenious twist to
the standard DLVO theory. In their theory, the Gibbs pair potential
acquires an attractive tail which can overcome the electrostatic repulsions
at large distances. Sogami and Ise argue that this is the explanation
for the void structures and other anomalies seen in colloidal
suspensions at low ionic strengths.

Of course, the actual forces must be independent of the thermodynamic
ensemble used:\cite{KjMi} 
the force experienced by $i$'th macroion at position
$\r_i$ is ${\mathbf{F}}_i = -\bigl({\partial
F}/{\partial\r_i}\bigr)_{V,T} = -\bigl({\partial
G}/{\partial\r_i}\bigr)_{p,T}$.  It is easy to dismiss the
SI theory therefore, on the grounds that an elementary mistake has
been made.  However, if it is claimed that the SI potential is only
an ``effective pair potential,''\cite{bi1} then what the SI theory
actually does is to capture the state point dependence of the Debye
screening length. This is most clearly illustrated in the derivation
by Schmitz.\cite{Schmitz2}  In fact, this model (charged particles
with pairwise screened Coloumb interactions and a state point
dependent screening length) \emph{has} been shown to have
vapour-liquid coexistence, in direct simulations by Dijkstra and van
Roij,\cite{DR} thus vindicating the SI attraction mechanism.

Such a model omits the volume terms described in the main text though,
and I would argue only captures a small part of the physics of the
real situation.  Nevertheless, one should credit Sogami and Ise with
identifying the essential electrostatic origin of the effective
attraction mechanism.  For instance: ``the attraction in question is a
logical consequence of the principle of electric
neutrality.''\cite{SI}

\section{Comparison with the van Roij and Hansen theory}
More recently, van Roij, Dijkstra and Hansen (RDH) have studied charge
stabilised colloids by density functional theory.\cite{vRH,vRDH} By
making what is essentially a \emph{random phase approximation} (RPA)
in the inhomogeneous density functional, they are able to solve the
problem analytically at least as far as computing a closed form
expression for the free energy. The resulting free energy is basically
the same as the present DH free energy, in the sense that for each
important term in Eq.~\eqref{ftoteq}, there is a corresponding term in
the RDH theory with more or less the same functional dependence. They
report phase diagrams for only a few parameters, which are compared
with the present theory in Table~\ref{tab:vRH}. (It should be noted
that the longer paper\cite{vRDH} corrects the initially reported
calculations.\cite{vRH}) The predictions from both theories for the
critical salt concentration are in good agreement, but the critical
volume fractions are as much as an order of magnitude higher in RDH at
the high $Z$ end.  At the highest $Z$, RDH do not appear to find even
metastable fluid-fluid coexistence.  The trends are not the same
either: in the present theory the critical $\phi$ increases with
increasing $Z$, whereas the reverse holds for RDH.  Similarly, the
present theory always predicts a lower critical salt concentration,
albeit at very low salt concentrations.  It is not clear if this also
holds for RDH.  These differences in detail though should not mask the
overall similarity between the predictions of the two theories.

\vspace{24pt}

\begin{table}
\begin{tabular}{rrccccc}
&&&\multicolumn{4}{c}{Upper critical solution state-point}\\
&&&\multicolumn{2}{c}{Present work}&\multicolumn{2}{c}{van Roij 
  \etal\ \cite{vRDH}}\\
$\sigma/\nm$&$Z$&$Z\lB/\sigma$&$\phi$&$\salt/\M$&$\phi$&$\salt/\M$\\
\tableline
 652& 7300& 8.06& $4.08\times10^{-3}$&$2.78\times10^{-5}$
&---&---\\
 461& 3650& 5.70& $6.29\times10^{-3}$&$3.10\times10^{-5}$
&$7.6\times10^{-2}$&$2.0\times10^{-5}$\\
 349& 2086& 4.30& $8.48\times10^{-3}$&$3.25\times10^{-5}$
&$3.8\times10^{-2}$&$2.1\times10^{-5}$\\
 266& 1217& 3.29& $9.81\times10^{-3}$&$3.25\times10^{-5}$
&$1.4\times10^{-2}$&$1.9\times10^{-5}$\\
\end{tabular}
\vspace{24pt}
\caption[?]{Upper critical solution points for van Roij, Dijkstra and
Hansen macroion parameters.\cite{vRDH}\label{tab:vRH}}
\end{table}

In the theory, the differences lie in three places.  Firstly, in the
macroion-small ion interaction term, RDH end up with an expression
very similar to the present one in Eq.~\eqref{fsmseq}, but with a
somewhat different finite size scaling factor: $f(x)=3/(4(1+x))$ (note
that $f(x)\not\to1$ as $x\to0$).  Which is closer to the true
free energy depends on whose approximations are most believable; for
example one can compare the difference to the two approximations
introduced for the MSA by Groh \etal.\cite{GED} Secondly there are minor
differences in macroion-small ion hard core interaction term---their
expression can be derived from the present theory by setting
$\log({1}/({1-\phi})) = \log(1+{\phi}/({1-\phi})) \approx
{\phi}/({1-\phi})$ and making a similar adjustment in the
prefactor. Thirdly, the small ion interaction term is absent from RDH
although it makes no practical difference to the results.  This term
can be recovered if one actually carries out the functional
integration over the small ion density fluctuations.\cite{SFE}

A point of fundamental interest remains though.  In their longer
paper,\cite{vRDH} RDH claim that the long wavelength
macroion-macroion structure factor, $\Scc(q\to0)$, should \emph{not}
diverge at the critical point because the pair interactions between
macroions remain purely repulsive, although in a later article they
modify this position somewhat.\cite{HGvR} The counter-argument is that
the spinodal line (including the critical point) is the locus of
points where the determinant of second partial derivatives of the free
energy vanishes. Since the long wavelength structure factor
\emph{matrix} is the inverse of the matrix of second partial
derivatives of the free energy, one would expect \emph{every}
component of the structure factor matrix including $\Scc(q\to0)$ to
diverge on the spinodal line, barring accidental cancellations.

What, then, is missing from the pair potential argument?  Firstly, we
have seen that many-body effects are undoubtably present, although
concealed in the present theoretical approaches by the apparent
simplicity of the electroneutrality constraint. Secondly, the use of
the RPA itself can be misleading. Frequently, the RPA is used to
derive an $S(q)$ which is at a \emph{lower} level of approximation
(zeroth order in a loop expansion to be precise) than the free energy
(first order in a loop expansion). If the phase instability only
appears at the higher level of approximation, as is the case for DH
theory for example, then it will not be seen in $S(q)$. This does not
uncover any profound physics though, rather it represents an
inconsistent level of approximation between $S(q)$ and the free
energy.  To get consistent results, one should either compute $S(q)$
to first order in a formal loop expansion (which may be a hard
calculation), or invert the matrix of second partial derivatives of
the free energy (which only gives $S(q\to0)$).  I have also emphasised
these considerations in relation to polyelectrolytes: see appendix
to.\cite{PBW}

\section{Comparison with the theory of Levin, Barbosa and Tamashiro.}
In a series of papers, Levin, Barbosa and Tamashiro (LBT) introduce a
theory for charge stabilised colloidal suspensions which is very close
to the present studies.\cite{LBT,LBT2} The theory is studied in the
absence of added salt though, and unfortunately this may have led the
authors to miss the existence of the miscibility gap which only
appears at a finite added salt concentration (according to the present
theory, at least).  A novel feature of the LBT approach is an attempt
to account for the counterion condensation phenomenon.  Their results
for the effective charge resemble the PB theory of Alexander
\etal.\cite{ACGMPH} and Groot.\cite{GrootIon} Thus Fig.~1 in
Ref.~\onlinecite{LBT} shows counterion condensation starting
effectively for $Z\lB/\sigma\approx3$ (their reported calculation is
for $Z\approx600$ and $a/\lB=\Tstar=100$, where $\Tstar$ is their
reduced temperature).

The LBT approach is motivated by analogy to the work on the RPA by
Fisher and Levin.  The free energy is split into contributions similar
to those identified here, and the incorporation of the counterion
condensation is motivated by the importance of Bjerrum pairing in the
RPM.  The counterion contribution to the free energy is that of an
OCP, which reduces to the first of Eqs.~\eqref{fsmseq} here in the
relevant limit where the ion concentration is $\ll1\,\mathrm{M}$.  The
macroion-counterion contribution is identical to that in RDH, except
that the self energy of the macroion, $Z^2\lB\kT/2a$, is also
included.  This is essential for the LBT theory where the effective
charge $Z$ is variable, whereas in theories where $Z$ is fixed
the self energy just shifts the macroion chemical potential.  The
macroion contribution is treated via a mean field-van der Waals
approximation omitting the background subtraction counterterm.  As
described in the main text, this omission can have quite a significant
effect on the location of the miscibility gap, and it would be quite
important to rectify this in any future work.

It seems that progress might be made by combining the variational
approach to the counterion condensation phenomenon advocated by LBT,
with the more careful treatment of the free energy of the whole system
considered in the present paper (or by RDH).  It certainly would be
interesting to see whether the miscibility gap appears when added
salt is included in such a theory, which can legimately be pushed to
much higher values of the bare macroion charge than have been
considered in the present study.


\begin{references}
\bibitem{zeff}{The best formal justification that one should not
correct $Z$ in the background subtraction counterterm is given by RDH
although the technical reasons remain somewhat obscure.\cite{vRDH} The
physical reason is that $Zn_m$ is the net background uniform
counter-charge density required to make the macroion subsystem
well-defined.  In this respect both RDH and the present
work differ from Shih and Stroud.\cite{SS}}
\bibitem{attrnote}{Experiments sometimes suggest that charged colloid
particles can be attracted to a like charged planar surface,\cite{MIK}
which might at least partly be due to a polarisation attraction due to
the ionic strength gradient (but see also Refs.~\onlinecite{QB} and
\onlinecite{AAL}).}
\end{references}
\end{document}